\documentclass[12pt]{iopart}
\usepackage{iopams}

\newtheorem{theorem}{Theorem}[section]
\newtheorem{proposition}[theorem]{Proposition}

\newlength{\vshift}
\newlength{\hshift}
\usepackage{amssymb,amsopn}

\begin{document}


\title[Characterization of $({\cal R},p,q)$-deformed Rogers-Szeg\"o polynomials]{Characterization of $({\cal R},p,q)-$deformed
 Rogers-Szeg\"o polynomials: associated quantum algebras, deformed Hermite polynomials
and relevant properties}
\author{J D Bukweli Kyemba and M N Hounkonnou}
\address{International Chair of Mathematical Physics
and Applications (ICMPA-UNESCO Chair), University of
Abomey-Calavi, 072 B.P.: 50 Cotonou, Republic of Benin}
\eads{\mailto{norbert.hounkonnou@cipma.uac.bj},
\mailto{hounkonnou@yahoo.fr}\\
\mailto{desbuk@gmail.com}}

\begin{abstract}
This paper addresses a new characterization of  $({\cal R},p,q)-$deformed Rogers-Szeg\"o
polynomials by providing their three-term recurrence relation and the associated quantum algebra built with  corresponding 
creation and annihilation operators. The whole construction is performed in  
 a unified way, generalizing all known relevant results which are straightforwardly derived as particular cases. Continuous $({\cal
R},p,q)-$deformed  Hermite polynomials and their recurrence
relation are also deduced. Novel relations are provided and discussed.
\end{abstract}

 \submitto{\JPA}
 \today



\section{Introduction}
\label{sect:intro}
Deformed quantum algebras, namely the $q-$ deformed algebras \cite{Jimbo,Odzij,QPT}
and their extensions to $(p,q)-$deformed algebras \cite{BurKlim,ChakJagan},
continue to attract much attention. One of the main reasons is that these topics represent a meeting point of nowadays fast developing areas
in mathematics and  physics like the theory of quantum orthogonal polynomials and special functions, quantum groups, integrable systems and quantum
and conformal field  theories and statistics.
Indeed, since the work of Jimbo \cite{Jimbo},  these fields have known profound interesting  developments which can be partially
found, for instance,
in the books by Chari  and  Pressley \cite{ChaPres},
Klimyk and Schumudgen \cite{KASK}, Ismail Moudard \cite{IMEH} and references therein.

The two-parameter quantum algebra, $U_{p,q}(gl(2))$, was
first introduced in reference \cite{ChakJagan} in view to
generalize or/and unify a series of  $q-$oscillator algebra variants,
 known in the earlier physics and mathematics
literature on the representation theory of
single-parameter quantum algebras. Then flourish investigations   in the same direction, among which
 the
work of  Burban and Klimyk \cite{BurKlim} on representations of
two-parameter quantum groups  and models of
two parameter quantum algebra $U_{p,q}(su_{1,1})$ and
$(p,q)-$deformed oscillator algebra. Almost simultaneously, Gelfand et al.\cite{GelfanGV}   introduced the
$(r,s)-$hypergeometric series  satisfying two parameter
difference equation, including $r-$ and $s-$shift operators. This new series
reproduces the Burban and Klimyk's $P,Q-$hypergeometric functions. The $(p,q)-$deformation rapidly found
 applications in
physics and mathematical physics as described for instance in
\cite{FLV,HmNe,HmNe2}.

Upon recalling a technique of constructing explicit realizations
of raising and lowering operators that satisfy an algebra akin to
the usual harmonic oscillator algebra, through the use of the
three-term recurrence relation and the differentiation expression
of Hermite polynomials, Galetti \cite{Galetti} has shown that a similar procedure
can be carried out in the case of  the three-term recurrence
relation for Rogers-Szeg\"o  and Stieltjes-Wigert polynomials and
the Jackson $q-$derivative. This technique furnished new realizations
of the $q-$deformed algebra associated with the $q-$deformed
harmonic oscillator, which obey,  well known and spread in the literature,  commutation
relations.

In the same vein, after recalling the  connection between the
Rogers-Szeg\"o polynomials and the $q-$oscillator, Jagannathan and
Sridhar \cite{JaganSridh} have defined a $(p,q)-$Rogers-Szeg\"o polynomials,  shown
that they are connected with the $(p,q)-$deformed oscillator
associated with the Jagannathan-Srinavasa $(p,q)-$numbers \cite{JaganRao} and
proposed a new realization of this algebra. In a previous
paper \cite{HounkBuk}, we have proposed a theoretical framework for the $(p,q)-$deformed state generalization
and
  provided a
generalized deformed quantum algebra, based  on a
work by Odzijewicz \cite{Odzij} on a  generalization of
$q-$deformed states in which the realizations of creation and
annihilation operators are given by multiplication by $z$ and the
action of the deformed derivative $\partial_{{\cal R},p,q}$ on the
space of analytic functions defined on the disc.

 The present
investigation aims at
giving a new realization of the previous
generalized deformed quantum algebras and an  explicit definition of
 the $({\cal R},p,q)-$Rogers-Szeg\"o polynomials, together with
their three-term recurrence relation and  the deformed difference
equation giving rise to the creation and annihilation operators.

The paper is organized as follows. As a matter of clarity, we
present in Section \ref{Sec2} a brief review of known results on $({\cal R},p,q)-$deformed
numbers, binomial coefficients and quantum algebra.
In section \ref{Sec3}, 
 we perform the realization of $({\cal
R},p,q)-$deformed quantum algebras using the $({\cal
R},p,q)-$difference equation and  the three-term recurrence
relation satisfied by $({\cal R},p,q)-$Rogers-Szeg\"o polynomials. The
key result of this section is the Theorem \ref{theoBuk}
giving the method of computation of relevant quantities. Section \ref{Sec4}
is devoted to the study of the  continuous $({\cal R},p,q)-$Hermite
polynomials. We then give their definition and  recurrence
relation. In Section \ref{Sec5}, relevant examples and their properties are provided and demonstrated.
Finally, Section \ref{Sec6} ends with the concluding remarks.

\section{$({\cal R},p,q)-$numbers and  associated $({\cal R},p,q)-$deformed
quantum algebras}\label{Sec2}
This section addresses the general theoretical framework as well as a brief review of known results on deformed
numbers and  deformed binomial coefficients.
The calculus methodology leading to   the definition and the computation
of the three-term recurrence relation of polynomials is also exposed.

In \cite{HounkBuk}, we have derived the $({\cal R},p,q)-$numbers
which are a generalization of Heine $q-$number
\begin{eqnarray}
[n]_q=\frac{1-q^n}{1-q}, \qquad n= 0, 1, 2, \cdots
\end{eqnarray}
and Jagannathan-Srinivasa $(p,q)-$numbers \cite{JaganRao},
\begin{eqnarray}\label{pqnbrs}
[n]_{p,q}= \frac{p^n-q^n}{p-q}, \qquad n= 0, 1, 2, \cdots.
\end{eqnarray}
Consider $p$ and $q$,  two
positive real numbers  such that $0<q<p$, and a given meromorphic function
${\cal R}$, defined on $\mathbb{C}\times\mathbb{C}$ by
\begin{equation}\label{Rxy}
{\cal R}(x,y) = \sum_{k, l = - L}^{\infty} r_{kl}x^ky^l
\end{equation}
with an eventual isolated singularity at the zero, where $r_{kl}$
are complex numbers, $L\in\mathbb{N}\cup\{0\}$, ${\cal R}(p^n, q^n)>
0$ $\forall n\in\mathbb{N}$, and ${\cal R}(1, 1) = 0$.
Denote by $\mathbb{D}_R$ =$\{z\in\mathbb{C}$ : $\vert z\vert < R \}$ a complex disc  and by
${\cal O}(\mathbb{D}_R)$ the set of holomorphic functions defined
on $\mathbb{D}_R$.
Then, the  $({\cal R},p,q)-$number is given by \cite{HounkBuk}
\begin{eqnarray}
[n]_{{\cal R},p,q}= {\cal R}(p^n,q^n),\qquad n= 0, 1, 2, \cdots
\end{eqnarray}
 leading to define $({\cal R},p,q)-$factorials
\begin{eqnarray}
[n]!_{{\cal R},p,q}= \left\{\begin{array}{lr} 1 \quad \mbox{for   } \quad n=0 \quad \\
{\cal R}(p,q)\cdots{\cal R}(p^n,q^n) \quad \mbox{for } \quad n\geq
1, \quad \end{array} \right.
\end{eqnarray}
and the  $({\cal R},p,q)-$binomial coefficients
\begin{eqnarray}
\left[\begin{array}{c} m  \\ n\end{array} \right]_{{\cal R},p,q}=
\frac{[m]!_{{\cal R},p,q}}{[n]!_{{\cal R},p,q}[m-n]!_{{\cal R},p,q}},\quad m, n= 0, 1, 2, \cdots;\quad m\geq n
\end{eqnarray}
that satisfy the relation
\begin{eqnarray}
 \left[\begin{array}{c} m  \\ n\end{array} \right]_{{\cal R},p,q}=
\left[\begin{array}{c} m  \\ m-n\end{array} \right]_{{\cal R},p,q},\quad m, n= 0, 1, 2, \cdots;\quad m\geq n.
\end{eqnarray}
There also result the following linear operators  defined on ${\cal O}(\mathbb{D}_R)$ by
(see \cite{HounkBuk} and references therein for more details):
\begin{eqnarray}\label{operat}
 &&\quad Q: \varphi \longmapsto Q\varphi(z) = \varphi(qz)
\nonumber \\
 &&\quad P: \varphi \longmapsto P\varphi(z) = \varphi(pz) \nonumber\\
&&{\label{deriva}} \partial_{p,q}:\varphi \longmapsto
\partial_{p,q}\varphi(z) = \frac{\varphi(pz) - \varphi(qz)}{z(p-q)},
\end{eqnarray}
and the $({\cal R},p,q)-$ derivative given by
\begin{equation}{\label{deriva1}}
\partial_{{\cal R},p,q} := \partial_{p,q}\frac{p - q}{P-Q}{\cal R}(P, Q)
= \frac{p - q}{pP-qQ}{\cal R}(pP, qQ)\partial_{p,q}.
\end{equation}

The quantum algebra associated with the $({\cal R},p,q)-$deformation,
denoted by ${\cal A}_{{\cal R},p,q}$, is generated by the
set of operators $\{1, A, A^\dag, N\}$ satisfying 
\begin{eqnarray}
&& \label{algN1}
\quad A A^\dag= [N+1]_{{\cal R},p,q},\quad\quad\quad A^\dag  A = [N]_{{\cal R},p,q};
\cr&&\left[N,\; A\right] = - A, \qquad\qquad\quad \left[N,\;A^\dag\right] = A^\dag
\end{eqnarray}
with the realization on ${\cal O}(\mathbb{D}_R)$ given by \cite{HounkBuk}
\begin{eqnarray}
A^\dag \equiv z,\qquad A\equiv\partial_{{\cal R},p,q}, \qquad N\equiv z\partial_z,
\end{eqnarray}
where $\partial_z\equiv\frac{\partial}{\partial z}$ is the usual derivative on $\mathbb{C}$.


\section{$({\cal R},p,q)-$ Rogers-Szeg\"o polynomials and their related quantum algebras}\label{Sec3}
This section aims at providing  realizations of $({\cal
R},p,q)-$deformed quantum algebras induced by $({\cal
R},p,q)-$Rogers-Szeg\"o polynomials. We first define the latter
and their three-term
recurrence relation, and then following the procedure elaborated in
\cite{Galetti, JaganSridh}, we prove that every sequence of these
polynomials forms a basis for the corresponding deformed quantum
algebra.

Indeed, Galetti in \cite{Galetti}, upon recalling the technique of
construction of raising and lowering operators which satisfy an
algebra akin to the usual harmonic oscillator algebra, by using
the three-term recurrence relation and the differentiation
expression of Hermite polynomials, has shown that a similar
procedure can be carried out to construct a
$q$-deformed harmonic oscillator algebra, with the help of relations
controlling the   Rogers-Szeg\"o polynomials. Following this author,
Jagannathan and Sridhar in \cite{JaganSridh} adapted the same approach
 to construct a Bargman-Fock
realization of the harmonic oscillator as well as realizations of
$q$- and  $(p,q)$-
deformed harmonic oscillators based on  Rogers-Szeg\"o polynomials.

As matter of clarity, this section is stratified as follows. We first develop  the synoptic schemes of
known different generalizations and then  display the formalism of
$({\cal R},p,q)$-Rogers-Szeg\"o polynomials.

\subsection{Hermite polynomials and harmonic oscillator approach}
The Hermite polynomials are defined as orthogonal polynomials  satisfying the
three-term recurrence relation
\begin{eqnarray}\label{Hermttr}
\mathbb{H}_{n+1}(z)=2z\mathbb{H}_n(z)-2n\mathbb{H}_{n-1}(z)
\end{eqnarray}
and the differentiation relation
\begin{eqnarray}\label{Hermdiff}
\frac{d}{dz}\mathbb{H}_n(z)= 2n\mathbb{H}_{n-1}(z).
\end{eqnarray}
Inserting Eq. (\ref{Hermdiff}) in Eq. (\ref{Hermttr}), one gets
\begin{eqnarray}
\mathbb{H}_{n+1}(z)= \left(2z-\frac{d}{dz}\right)\mathbb{H}_n(z)
\end{eqnarray}
which includes the introduction of a raising operator (see
\cite{Galetti} and references therein), defined as
\begin{eqnarray}
\hat{a}_+= 2z-\frac{d}{dz}
\end{eqnarray}
such that the set of Hermite polynomials can be generated by the
application of this operator to the first polynomial
$\mathbb{H}_0(z)=1$, i.e.,
\begin{eqnarray}
\mathbb{H}_n(z)= \hat{a}_+^n \mathbb{H}_0(z).
\end{eqnarray}
From Eq.(\ref{Hermdiff}), one defines the lowering operator $\hat{a}_-$ as
\begin{eqnarray}
\hat{a}_-\mathbb{H}_n(z)= \frac{1}{2}\frac{d}{dz}\mathbb{H}_n(z)=n\mathbb{H}_{n-1}(z).
\end{eqnarray}
Furthermore one constructs a number operator in the form
\begin{eqnarray}
\hat{n}= \hat{a}_+\hat{a}_-.
\end{eqnarray}
One can readily check that these operators satisfy the canonical commutation relations
\begin{eqnarray}\label{HarmAlg}
[ \hat{a}_-,\;\hat{a}_+]=1,\qquad [ \hat{n},\;\hat{a}_-]=-\hat{a}_- ,\qquad [ \hat{n},\;\hat{a}_+]=\hat{a}_+,
\end{eqnarray}
although the operators $\hat{a}_-$  and $\hat{a}_+$ are not the
usual creation and annihilation  operators associated
with the quantum mechanics harmonic oscillator. Thus, we see that
one can obtain  raising,  lowering and  number operators from
the two basic relations satisfied by the Hermite polynomials,
i. e. the three-term recurrence relation and the differentiation
relation, respectively, so that they satisfy the well known
commutation relations.

On the other hand, if one considers the usual Hilbert space
spanned by the vectors $|n\rangle$, generated from the vacuum
$|0\rangle$ by the raising operator $\hat{a}_+$, then together
with the lowering operator $\hat{a}_-$, the following relations
hold
\begin{eqnarray}
&&\hat{a}_-\hat{a}_+-\hat{a}_+\hat{a}_+= 1,\cr
&&\langle 0|0\rangle =1,\cr
&&|n\rangle = \hat{a}_+^n|0\rangle,\cr
&&\hat{a}_-|0\rangle =0.
\end{eqnarray}
In particular,  the next expressions, established using  the
previous equations, are in order:
\begin{eqnarray}
&&\hat{a}_+|n\rangle = |n+1\rangle,\cr
&&\hat{a}_-|n\rangle = |n-1\rangle,\cr
&&\langle m|n\rangle =n!\delta_{mn}.
\end{eqnarray}

Now, on the other hand,  examining  the procedure given in \cite{JaganSridh}, the authors considered the  sequence of polynomials
\begin{eqnarray}
\psi_n(z)=\frac{1}{\sqrt{n!}}{\bf h}_n(z),
\end{eqnarray}
where
\begin{eqnarray}
{\bf h}_n(z)=(1+z)^n =\sum_{k=0}^{n}\left(\begin{array}{c}n\\k\end{array}\right) z^k,
\end{eqnarray}
obeying the  relations
\begin{eqnarray}
\frac{d}{dz}\psi_n(z)&=&\sqrt{n}\psi_{n-1}(z),\\
(1+z)\psi_n(z) &=&\sqrt{n+1} \psi_{n+1}(z),\label{Hermttr1}\\
(1+z)\frac{d}{dz}\psi_n(z)&=&n\psi_{n}(z),\label{Hermdiff1}\\
\frac{d}{dz}\left((1+z)\psi_n(z)\right)&=&(n+1)\psi_n(z).
\end{eqnarray}
Here equations (\ref{Hermttr1}) and (\ref{Hermdiff1}) are the recurrence relation
and the differential equation for polynomials $\psi_n(z),$ respectively.
By analogy to the work done by Galleti, Jagannathan and Sridhar proposed the following relations:
\begin{eqnarray}
\hat{a}_+=(1+z),\qquad \hat{a}_-=\frac{d}{dz}, \qquad \hat{n}= (1+z)\frac{d}{dz},
\end{eqnarray}
for creation (or raising), annihilation (or lowering) and number
operators, respectively, and found that  the set
$\{\psi_n(z)\;|n=0, 1, 2, \cdots\}$ forms a basis for the
Bargman-Fock realization of the harmonic oscillator
(\ref{HarmAlg}).

\subsection{Rogers-Szeg\"o polynomials and $q$-deformed harmonic oscillator}\label{qHosc}
Here in analogous way as  Jagannathan and
Sridhar \cite{JaganSridh}, we perform a  construction of the creation, annihilation and
number operators from the three-term recurrence relation and the
$q-$difference equation founding the Rogers-Szeg\"o
polynomials. This procedure  a little differs from
 that used by Galetti \cite{Galetti} to obtain raising, lowering and number operators.\\
The Rogers-Szeg\"o polynomials are defined as
\begin{eqnarray}
H_n(z;q)=\sum_{k=0}^{n}\left[\begin{array}{c}n\\k\end{array}\right]_q z^k,\quad n=0,1,2\cdots
\end{eqnarray}
and satisfy a three-term recurrence relation
\begin{eqnarray}\label{qRogersttr}
H_{n+1}(z;q)=(1+z)H_n(z;q)-z(1-q^n)H_{n-1}(z;q)
\end{eqnarray}
as well as the $q$-difference equation
\begin{eqnarray}
\partial_qH_n(z;q)= [n]_qH_{n-1}(z;q).
\end{eqnarray}
In the limit case $q\to 1$, the Rogers-Szeg\"o polynomial of degree $n$ ($n= 0, 1, 2, \cdots$) well converges to
\begin{eqnarray*}
{\bf h}_n(z) =\sum_{k=0}^{n}\left(\begin{array}{c}n\\k\end{array}\right) z^k
\end{eqnarray*}
as required.
Defining
\begin{eqnarray}\label{qRogersnorm}
\psi_n(z;q)=\frac{1}{\sqrt{[n]!_q}}H_n(z)
=\frac{1}{\sqrt{[n]!_q}}\sum_{k=0}^{n}\left[\begin{array}{c}n\\k\end{array}\right]_q z^k,\quad n=0,1,2\cdots,
\end{eqnarray}
one can straightforwardly infer  that
\begin{eqnarray}\label{qRogersdiff}
\partial_q\psi_n(z;q)= \sqrt{[n]_q}\psi_{n-1}(z;q)
\end{eqnarray}
with the property that for $n=0, 1, 2, \cdots$
\begin{eqnarray}
\partial_q^{n+1}\psi_n(z;q) = 0\quad\mbox{and}\quad \partial_q^{m}\psi_n(z;q) \neq 0 \quad\mbox{for any  }
m<n+1.
\end{eqnarray}
It follows from Eqs. (\ref{qRogersttr}) and (\ref{qRogersnorm})
that the polynomials $\{\psi_n(z;q)\;|\; n= 0, 1, 2, \cdots\}$
satisfy the following three-term recurrence relation
\begin{eqnarray}\label{qRogersttr1}
 \sqrt{[n+1]_q}\psi_{n+1}(z;q)=(1+z)\psi_n(z;q)-z(1-q) \sqrt{[n]_q}\psi_{n-1}(z;q)
\end{eqnarray}
and the  $q-$difference equation
\begin{eqnarray}\label{qRogersttr2}
\left((1+z)-(1-q)z\;\partial_q\right)\psi_n(z;q)=  \sqrt{[n+1]_q}\psi_{n+1}(z;q)
\end{eqnarray}
obtained from Eq.(\ref{qRogersdiff}).
Hence, it is natural to formally define the number operator $N$  as
\begin{eqnarray}\label{qRogersnb}
N\psi_n(z;q)= n\psi_{n}(z;q)
\end{eqnarray}
determined for the creation and  annihilation operators expressed as
\begin{eqnarray}
A^\dag = 1+z-(1-q)z\;\partial_q \quad\mbox{and}\quad A = \partial_q
\end{eqnarray}
respectively.
Indeed, the proofs of the following  relations are immediate:
\begin{eqnarray}
N\psi_n(z;q)&=& n\psi_{n}(z;q),\\
A^\dag\psi_n(z;q)&=&  \sqrt{[n+1]_q}\psi_{n+1}(z;q),\\
A\psi_n(z;q)&=&  \sqrt{[n]_q}\psi_{n-1}(z;q),\\
A^\dag A\psi_n(z;q)&=&  [n]_q\psi_{n}(z;q) = [N]_q\psi_{n}(z;q),\\
AA^\dag\psi_n(z;q)&=&  [n+1]_q\psi_{n}(z;q) = [N + 1]_q\psi_{n}(z;q).
\end{eqnarray}
Therefore, one concludes that the set of polynomials $\{\psi_n(z;q)\;|n = 0, 1, 2,
\cdots \}$ provides a basis for a realization of the $q$-deformed
harmonic oscillator algebra given by
\begin{eqnarray}\label{qHarm}
AA^\dag - qA^\dag A=1,\qquad [ N, \;A]=-A ,\qquad [N,\;A^\dag]=A^\dag
\end{eqnarray}

\subsection{$({\cal R},p,q)-$generalized Rogers-Szeg\"o polynomials and quantum algebras}
We can now supply the general procedure for constructing
the recurrence relation for the $({\cal R},p,q)-$Rogers-Szeg\"o
polynomials and the related $({\cal R},p,q)$-difference
equation that allow to define the
creation, annihilation and number operators for a given $({\cal
R},p,q)-$deformed quantum algebra. This is summarized as follows.
\begin{theorem}\label{theoBuk}
If $\phi_i(x,y)$ ($i=1,2,3$) are functions satisfying:
\begin{eqnarray}
&&\phi_i(p,q)\neq 0\qquad \mbox{for   } i=1, 2, 3,\label{RpqRog0}\\
&&\phi_i(P,Q)z^k= \phi_i^k(p,q)z^k \quad \mbox{for }\; z\in\mathbb{C},\;k= 0, 1, 2,
\cdots\quad i=1, 2\label{RpqRog1}
\end{eqnarray}
and if, moreover, the following relation between $({\cal R},p,q)-$binomial coefficients holds
\begin{eqnarray}\label{RpqRog2}
&& \left[\begin{array}{c} n+1 \\ k \end{array}\right]_{{\cal R},p,q} =
\phi_1^k(p,q)\left[\begin{array}{c} n \\ k \end{array}\right]_{{\cal R},p,q} +
\phi_2^{n+1-k}(p,q)\left[\begin{array}{c} n \\ k-1 \end{array}\right]_{{\cal R},p,q}
\cr&&\qquad\qquad\qquad\qquad\qquad\qquad\qquad
-\phi_3(p,q)[n]_{{\cal R},p,q}\left[\begin{array}{c} n-1 \\ k-1 \end{array}\right]_{{\cal R},p,q}
\end{eqnarray}
for $1\leq k\leq n$, then the $({\cal R},p,q)-$Rogers-Szeg\"o polynomials defined as
\begin{eqnarray}
H_n(z;{\cal R},p,q)=\sum_{k=0}^{n}\left[\begin{array}{c}n\\k\end{array}\right]_{{\cal R},p,q} z^k,\quad n=0,1,2\cdots
\end{eqnarray}
satisfy the three-term recurrence relation
\begin{eqnarray}\label{RpqRogersttr}
H_{n+1}(z;{\cal R},p,q)&=& H_n\left(\phi_1(p,q)z:{\cal R},p,q\right)
\cr&&\qquad+ z\phi_2^n(p,q)H_n\left(z\phi_2^{-1}(p,q);{\cal R},p,q\right)
\cr&&\qquad-z\phi_3(p,q)[n]_{{\cal R},p,q}H_{n-1}\left(z;{\cal R},p,q\right)
\end{eqnarray}
and $({\cal R},p,q)-$difference equation
\begin{eqnarray}\label{RpqRogerdiff}
\partial_{{\cal R},p,q}H_n(z;{\cal R},p,q)= [n]_{{\cal R},p,q}H_{n-1}(z;{\cal R},p,q).
\end{eqnarray}
\end{theorem}
{\bf Proof:} Multiplying the two sides of the relation (\ref{RpqRog2}) by $z^k$ and adding for $k= 1$ to $n$ we get
\begin{eqnarray}
&&\sum_{k=1}^n\left[\begin{array}{c} n+1 \\ k \end{array}\right]_{{\cal R},p,q}z^k =
\sum_{k=1}^n\phi_1^k(p,q)\left[\begin{array}{c} n \\ k \end{array}\right]_{{\cal R},p,q}z^k
\cr&&\qquad\qquad\qquad\qquad\qquad +
\sum_{k=1}^n\phi_2^{n+1-k}(p,q)\left[\begin{array}{c} n \\ k-1 \end{array}\right]_{{\cal R},p,q}z^k
\cr&&\qquad\qquad\qquad\qquad\qquad
-\phi_3(p,q)[n]_{{\cal R},p,q}\sum_{k=1}^n\left[\begin{array}{c} n-1 \\ k-1 \end{array}\right]_{{\cal R},p,q}z^k.
\end{eqnarray}
After a short computation and using the condition (\ref{RpqRog2})
we get Eq.(\ref{RpqRogersttr}). Then there immediately results the  proof of
Eq.(\ref{RpqRogerdiff}).
\hfill $\square$\\
Setting
\begin{equation}
 \psi_{n}(z;{\cal R},p,q)= \frac{1}{\sqrt{[n]!_{{\cal R},p,q}}}H_n(z;{\cal R},p,q),
\end{equation}
and using the equations (\ref{RpqRogersttr}) and
(\ref{RpqRogerdiff}) yield the three-term recurrence relation
\begin{eqnarray}\label{RpqRogersttr1}
&&\left(\phi_1(P,Q)+ z\phi_2^n(p,q)\phi_2^{-1}(P,Q)
-z\phi_3(p,q)\partial_{{\cal R},p,q}\right)\psi_n(z;{\cal R},p,q)=\cr
&&\qquad\qquad\qquad\sqrt{[n+1]_{{\cal R},p,q}}\;\;\psi_{n+1}(z;{\cal R},p,q)
\end{eqnarray}
and $({\cal R},p,q)-$difference equation
\begin{eqnarray}\label{RpqRogerdiff1}
\partial_{{\cal R},p,q}\psi_n(z;{\cal R},p,q)= \sqrt{[n]_{{\cal R},p,q}}\;\;\psi_{n-1}(z;{\cal R},p,q)
\end{eqnarray}
for the polynomials $\psi_n(z;{\cal R},p,q)$ with the virtue that  for $n=0, 1, 2, \cdots$
\begin{eqnarray}
\partial_{{\cal R},p,q}^{n+1}\psi_n(z;{\cal R},p,q) = 0\;\mbox{and}\;\;
\partial_{{\cal R},p,q}^{m}\psi_n(z;{\cal R},p,q) \neq 0 \;\;\;\mbox{for }m<n+1.
\end{eqnarray}
Now, formally defining  the number operator $N$ as
\begin{eqnarray}
 N\psi_n(z;{\cal R},p,q)= n\psi_n(z;{\cal R},p,q),
\end{eqnarray}
and the raising  and lowering  operators by
\begin{eqnarray}
&&A^\dag =\left(\phi_1(P,Q)+ z\phi_2^N(p,q)\phi_2^{-1}(P,Q)
-z\phi_3(p,q)\partial_{{\cal R},p,q}\right)
 \;\mbox{and}\;
 \cr &&A = \partial_{{\cal R},p,q},
\end{eqnarray}
respectively, the set of  polynomials $\{\psi_n(z;{\cal
R},p,q)\;|\;n= 0, 1, 2, \cdots \}$ provides a basis for a
realization of $({\cal R},p,q)-$deformed quantum algebra ${\cal
A}_{{\cal R},p,q}$ satisfying the commutation relations
(\ref{algN1}). Provided the above formulated theorem, we can now show how the  realizations in
terms of Rogers-Szeg\"o polynomials  can be derived for different known deformations simply by
determining the functions $\phi_i$ ($i=1, 2, 3$) that satisfy the
relations (\ref{RpqRog0})-(\ref{RpqRog2}).

\section{Continuous $({\cal R}, p,q)-$Hermite polynomials}\label{Sec4}
We exploit here the  peculiar relation established in the   theory of $q-$deformation between Rogers-Szeg\"o polynomials and Hermite
polynomials \cite{IMEH,JaganRao,KASK,ASK}  and given by
\begin{eqnarray}
\mathbb{H}_n(\cos\theta;q)= e^{in\;\theta}H_n(e^{-2i\;\theta};q)=
\sum_{k=0}^n\left[\begin{array}{c}n\\k\end{array}\right]_q\;e^{i(n-2k)\theta}, n=0, 1, 2, \cdots,
\end{eqnarray}
where $\mathbb{H}_n$ and  $H_n$ stand for the Hermite and
Rogers-Szeg\"o polynomials, respectively. Is also of interest the
property that all the $q-$Hermite polynomials  can  be explicitly
recovered from the initial one $\mathbb{H}_{0}(\cos\theta;q)= 1$, using
the three-term recurrence relation
\begin{eqnarray}\label{qHttr}
 \mathbb{H}_{n+1}(\cos\theta;q)= 2\cos\theta\mathbb{H}_{n}(\cos\theta;q)-(1-q^{n})\mathbb{H}_{n-1}(\cos\theta;q)
\end{eqnarray}
with $\mathbb{H}_{-1}(\cos\theta;q)=0$.

In the same way we define the $({\cal R}, p,q)-$Hermite polynomials through the $({\cal R}, p,q)-$Rogers-Szeg\"o
polynomials as
\begin{eqnarray}
\mathbb{H}_n(\cos\theta;{\cal R}, p,q)= e^{in\;\theta}H_n(e^{-2i\;\theta};{\cal R}, p,q),\quad n=0, 1, 2, \cdots.
\end{eqnarray}
Then the next statement is true.
\begin{proposition}\label{propBuk}
Under the hypotheses of the theorem \ref{theoBuk}, the continuous $({\cal R}, p,q)$-Hermite polynomials satisfy
the following three-term recurrence relation
\begin{eqnarray}
\mathbb{H}_{n+1}(\cos\theta;{\cal R}, p,q)&=&
e^{i\;\theta}\phi_1^{\frac{n}{2}}(p,q)\phi_1(P,Q)\mathbb{H}_{n}(\cos\theta;{\cal R}, p,q)
\cr&&\quad + e^{-i\;\theta}\phi_2^{\frac{n}{2}}(p,q)\phi_2^{-1}(P,Q) \mathbb{H}_{n}(\cos\theta;{\cal R}, p,q)
\cr&&\quad-\phi_3(p,q)[n]_{{\cal R}, p,q}\mathbb{H}_{n-1}(\cos\theta;{\cal R}, p,q).\qquad
\end{eqnarray}
\end{proposition}

{\bf Proof:} Multiplying the two sides of the three-term
recurrence relation (\ref{RpqRogersttr}) by $e^{i(n+1)\theta},$ we
obtain, for $z= e^{-2i\theta}$,
\begin{eqnarray*}
e^{i(n+1)\theta}H_{n+1}(e^{-2i\theta};{\cal R},p,q)&=&
e^{i(n+1)\theta}H_n\left(\phi_1(p,q)e^{-2i\theta};{\cal R},p,q\right)
\cr&+& e^{i(n-1)\theta}\phi_2^n(p,q)H_n\left(\phi_2^{-1}(p,q)e^{-2i\theta};{\cal R},p,q\right)
\cr&-&e^{i(n-1)\theta}\phi_3(p,q)[n]_{{\cal R},p,q}H_{n-1}\left(e^{-2i\theta};{\cal R},p,q\right)
\cr&=&e^{i\theta}e^{in\theta}\phi_1(P,Q)H_n\left(e^{-2i\theta};{\cal R},p,q\right)
\cr&+& e^{-i\theta}\phi_2^n(p,q)e^{in\theta} \phi_2^{-1}(P,Q)H_n\left(e^{-2i\theta};{\cal R},p,q\right)
\cr&-& \phi_3(p,q)[n]_{{\cal R},p,q}e^{i(n-1)\theta}H_{n-1}\left(e^{-2i\theta};{\cal R},p,q\right).
\end{eqnarray*}
The required result follows  from the use of the equalities
\begin{eqnarray}
&&e^{in\theta}\phi_1(P,Q)H_n\left(e^{-2i\theta};{\cal
R},p,q\right)=
\phi_1^{\frac{n}{2}}(p,q)\phi_1(P,Q)e^{in\theta}H_n\left(e^{-2i\theta};{\cal
R},p,q\right), 
\cr&&e^{in\theta}\phi_2^{-1}(P,Q)H_n\left(e^{-2i\theta};{\cal R},p,q\right)=
\cr&&\qquad\qquad\qquad\qquad\qquad\phi_2^{-\frac{n}{2}}(P,Q)\phi_2^{-1}(P,Q)e^{in\theta}
H_n\left(e^{-2i\theta};{\cal R},p,q\right)
\end{eqnarray}
with
\begin{eqnarray}
 \phi_j(P,Q)e^{-2ik\theta} = \phi_j^k(p,q)e^{-2ik\theta}, \quad j=1,\;2,\; k= 0, 1, 2, \cdots.
\end{eqnarray}
\hfill $\square$\\


\section{Relevant particular cases}\label{Sec5}
The following pertinent cases deserve to be raised, as their derivation from the previous general theory
appeals concrete expressions for the deformed function ${\cal R}(p,q).$

\subsection{${\cal R}(x,y) = \frac{x-y}{p-q}$}\label{janasec}
%
In this case, the $({\cal R}, p,q)-$numbers are simply given by
\begin{eqnarray*}
 [n]_{p,q}={\cal R}(p^n,q^n)=\frac{p^n-q^n}{p-q},\quad n= 0,\; 1,\; 2,\; \cdots
\end{eqnarray*}
 with the $({\cal R}, p,q)-$factorials defined by
\begin{eqnarray}
[n]!_{p,q}= \left\{\begin{array}{lr} 1 \quad \mbox{for   } \quad n=0 \quad \\
\displaystyle\prod_{k=1}^n\frac{p^k-q^k}{p-q}=\frac{((p,q);(p,q))_n}{(p-q)^n} \quad \mbox{for } \quad n\geq
1. \quad \end{array} \right.
\end{eqnarray}
They correspond to the Jagannathan-Srinivasa $(p,q)$-numbers  and $(p,q)$-factorials\cite{JaganRao, JaganSridh}.

There result the following  relevant properties.
\begin{proposition}\label{PropJan1}
If $n$ and $m$ are nonnegative integers, then
\begin{eqnarray}
[n]_{p,q}&=& \sum_{k=0}^{n-1}p^{n-1-k}q^k,
\cr [n+m]_{p,q}&=& q^m[n]_{p,q}+p^n[m]_{p,q}= p^m[n]_{p,q}+q^n[m]_{p,q},
\cr [-m]_{p,q}&=& -q^{-m}p^{-m}[m]_{p,q},
\cr [n-m]_{p,q}&=& q^{-m}[n]_{p,q}-q^{-m}p^{n-m}[m]_{p,q}= p^{-m}[n]_{p,q}-q^{n-m}p^{-m}[m]_{p,q},\quad
\cr [n]_{p,q}&=& [2]_{p,q}[n-1]_{p,q}-pq[n-2]_{p,q}.
\end{eqnarray}
\end{proposition}

\begin{proposition}\label{PropJan2}
The $(p,q)-$binomial coefficients
\begin{eqnarray}
 \left[\begin{array}{c} n \\ k \end{array}\right]_{p,q}\equiv\frac{[n]!_{p,q}}{[k]!_{p,q}[n-k]!_{p,q}}=
\frac{((p,q);(p,q))_n}{((p,q);(p,q))_k((p,q);(p,q))_{n-k}}, 
\end{eqnarray}
where $0\leq k\leq n,\;\; n\in\mathbb{N},$ and $((p,q);(p,q))_m = (p-q)(p^2-q^2)\cdots(p^m-q^m)$,
$m\in\mathbb{N},$ satisfy the following identities:
\begin{eqnarray}
 \left[\begin{array}{c} n \\ k \end{array}\right]_{p,q}&=& \left[\begin{array}{c} n \\ n-k \end{array}\right]_{p,q}=
p^{k(n-k)}\left[\begin{array}{c} n \\ k \end{array}\right]_{q/p}=
p^{k(n-k)}\left[\begin{array}{c} n \\ n-k \end{array}\right]_{q/p},\label{Janeq1}
\\ \left[\begin{array}{c} n+1 \\ k \end{array}\right]_{p,q} &=& p^k\left[\begin{array}{c} n \\ k \end{array}\right]_{p,q}
+q^{n+1-k}\left[\begin{array}{c} n \\ k-1 \end{array}\right]_{p,q},\label{Janeq2}
\\ \left[\begin{array}{c} n+1 \\ k \end{array}\right]_{p,q} &=&
p^{k}\left[\begin{array}{c} n \\ k \end{array}\right]_{p,q} +
p^{n+1-k}\left[\begin{array}{c} n \\ k-1 \end{array}\right]_{p,q} -(p^n-q^n)
\left[\begin{array}{c} n-1 \\ k-1 \end{array}\right]_{p,q}\quad\label{Janeq3}
\end{eqnarray}
with
\begin{eqnarray*}
\left[\begin{array}{c} n \\ k \end{array}\right]_{q/p}= \frac{(q/p; q/p)_n}{(q/p; q/p)_k(q/p; q/p)_{n-k}},
\end{eqnarray*}
where $(q/p; q/p)_n = (1-q/p)(1-q^2/p^2)\cdots (1-q^n/p^n)$
and the $(p,q)-$shifted factorial
\begin{eqnarray}
((a,b);(p,q))_n &\equiv& (a-b)(ap-bq)\cdots(ap^{n-1}-bq^{n-1})\cr
 &=& \sum_{k=0}^{n}\left[\begin{array}{c} n \\ k \end{array}\right]_{p,q}(-1)^kp^{(n-k)(n-k-1)/2}
q^{k(k-1)/2}a^{n-k}b^k.
\end{eqnarray}
\end{proposition}

The algebra ${\cal A}_{p,q}$, generated by $\{1,\; A,\; A^\dag,\;N\}$, associated with
 $(p,q)-$Janagathan-Srinivasa deformation, satisfies the following commutation relations\cite{JaganRao, JaganSridh}:
\begin{eqnarray}
 A\;A^\dag- pA^\dag A= q^N, \quad &&A\;A^\dag- qA^\dag A= p^N\cr
[N,\;A^\dag]= A^\dag,\quad\qquad\quad&& [N,\;A]= -A.\label{JanSrialg}
\end{eqnarray}

\label{pqHosc}
The $(p,q)$-Rogers-Szeg\"o polynomials  studied in \cite{JaganSridh}
appear as a particular case obtained by choosing
$\phi_1(x,y)=\phi_2(x,y)=\phi(x,y)= x$ and $\phi_3(x,y)= x-y$. Indeed,\\
$\phi(p,q)= p\neq 0$, $ \phi_3(p,q)=p-q\neq 0$, $\phi(P,Q)z^k=
\phi_1^k(p,q)z^k$ and Eq.(\ref{Janeq3}) shows that
$$\left[\begin{array}{c} n+1 \\ k \end{array}\right]_{p,q} =
p^{k}\left[\begin{array}{c} n \\ k \end{array}\right]_{p,q} +
p^{n+1-k}\left[\begin{array}{c} n \\ k-1 \end{array}\right]_{p,q} -(p-q)[n]_{p,q}
\left[\begin{array}{c} n-1 \\ k-1 \end{array}\right]_{p,q}.$$
Hence, the hypotheses of the above theorem are satisfied and, therefore,
the $(p,q)-$Rogers-Szeg\"o polynomials
\begin{eqnarray}
 H_n(z;p,q)=\sum_{k=0}^{n}\left[\begin{array}{c}n\\k\end{array}\right]_{p,q}z^k\quad n= 0, 1, 2, \cdots
\end{eqnarray}
satisfy the three-term recurrence relation
\begin{eqnarray}\label{JannRogersttr}
H_{n+1}(z;p,q)&=& H_n(pz;p,q) + z p^nH_n(p^{-1}z;p,q)\cr&&-z(p^n-q^n)H_{n-1}(z;p,q)
\end{eqnarray}
and $(p,q)-$difference equation
\begin{eqnarray}\label{JanRogerdiff}
\partial_{p,q}H_n(z;p,q)= [n]_{p,q}H_{n-1}(z;p,q).
\end{eqnarray}
Finally, the set of polynomials
\begin{eqnarray}
 \psi_n(z;p,q)= \frac{1}{\sqrt{[n]!_{p,q}}}H_n(z;p,q),\quad n=0, 1, 2, \cdots
\end{eqnarray}
forms a basis for a realization of the $(p,q)-$deformed harmonic
oscillator and quantum algebra ${\cal A}_{p,q}$ satisfying
the commutation relations (\ref{JanSrialg}) with the number
operator $N$  defined  as
\begin{eqnarray}
 N\psi_n(z;p,q)= n\psi_n(z;p,q),
\end{eqnarray}
relating  the annihilation and  creation operators  given by
\begin{eqnarray}
 A= \partial_{p,q}\quad\mbox{and}\quad A^\dag= P+ zp^NP^{-1}-z(p-q)\partial_{p,q}
\end{eqnarray}
respectively. Naturally, setting $p=1$ one recovers the results of the
subsection~\ref{qHosc}.

The continuous $(p,q)-$Hermite polynomials have been already suggested in
\cite{JaganRao} without any further details. In the above achieved
generalization, these polynomials are given by
\begin{eqnarray}
 \mathbb{H}_n(\cos\theta;p,q)&=& e^{in\theta}H_n(e^{-2i\theta};p,q)\cr&=&
\sum_{k=0}^n\left[\begin{array}{c}n\\k\end{array}\right]_{p,q}\;e^{i(n-2k)\theta},\quad n=0, 1, 2, \cdots.
\end{eqnarray}
Since for the $(p,q)-$deformation  $\phi_1(x,y)=\phi_2(x,y)= x$ and
$\phi_3(x,y)=x-y$, from the Proposition \ref{propBuk} we deduce
that the corresponding sequence of continuous $(p,q)-$polynomials
satisfies the three-term recurrence relation
\begin{eqnarray}\label{pqHttr}
 \mathbb{H}_{n+1}(\cos\theta;p,q)&=& p^{\frac{n}{2}}(e^{i\theta}P+e^{-i\theta}P^{-1})\mathbb{H}_n(\cos\theta;p,q)
\cr&&-(p^n-q^n)\mathbb{H}_{n-1}(\cos\theta;p,q),
\end{eqnarray}
with $Pe^{i\theta}=p^{-1/2}e^{i\theta}$. This relation turns to be the well-known
three-term recurrence relation (\ref{qHttr}) for continuous $q-$Hermite polynomials in the limit $p\to 1$.
As matter of illustration, let us  explicitly compute the first three  polynomials  using the relation (\ref{pqHttr}), with
$\mathbb{H}_{-1}(\cos\theta;p,q)=0$ and $\mathbb{H}_0(\cos\theta;p,q)=1$:\
\begin{eqnarray*}
\mathbb{H}_1(\cos\theta;p,q)&=&p^0(e^{i\theta}P+e^{-i\theta}P^{-1})1-(p^0-q^0)0 = e^{i\theta}+e^{-i\theta}=2\cos\theta\cr
&=&\left[\begin{array}{c}1\\0\end{array}\right]_{p,q}e^{i\theta}+\left[\begin{array}{c}1\\1\end{array}\right]_{p,q}e^{-i\theta}.\cr
\mathbb{H}_2(\cos\theta;p,q)&=& p^{\frac{1}{2}}(e^{i\theta}P+e^{-i\theta}P^{-1})(e^{i\theta}+e^{-i\theta})
-(p-q)1  \cr&=& e^{2i\theta}+e^{-2i\theta}+p+q= 2\cos2\theta +p+q
 \cr&=&\left[\begin{array}{c}2\\0\end{array}\right]_{p,q}e^{2i\theta}+\left[\begin{array}{c}2\\1\end{array}\right]_{p,q}e^{0i\theta}
+\left[\begin{array}{c}2\\2\end{array}\right]_{p,q}e^{-i\theta}.
\end{eqnarray*}
\begin{eqnarray*}
&&\mathbb{H}_3(\cos\theta;p,q)= p(e^{i\theta}P+e^{-i\theta}P^{-1})(e^{2i\theta}+e^{-2i\theta}+p+q)
\cr&&\qquad\qquad\qquad\qquad\qquad\qquad\quad-(p^2-q^2)(e^{i\theta}+e^{-i\theta})
\cr&&\qquad\qquad= e^{3i\theta}+e^{-3i\theta}+(p^2+pq+q^2)(e^{i\theta}+e^{-i\theta})
\cr&&\qquad\qquad= 2\cos3\theta+2(p^2+pq+q^2)\cos\theta\cr
&&\qquad\qquad=\left[\begin{array}{c}3\\0\end{array}\right]_{p,q}e^{3i\theta}+\left[\begin{array}{c}3\\1\end{array}\right]_{p,q}e^{i\theta}
+\left[\begin{array}{c}3\\2\end{array}\right]_{p,q}e^{-i\theta}+\left[\begin{array}{c}3\\3\end{array}\right]_{p,q}e^{-3i\theta}.
\end{eqnarray*}

\subsection{${\cal R}(x,y)= \frac{1-xy}{(p^{-1}-q)x}$}
%
The $({\cal R},p,q)$-numbers and $({\cal R},p,q)-$factorials are reduced to $(p^{-1},q)$-numbers and
$(p^{-1},q)-$factorials, namely,
\begin{eqnarray*}
 [n]_{p^{-1},q}=\frac{p^{-n}-q^n}{p^{-1}-q},
\end{eqnarray*}
and
\begin{eqnarray}
[n]!_{p^{-1},q}= \left\{\begin{array}{lr} 1 \quad \mbox{for   } \quad n=0 \quad \\
\frac{((p^{-1},q);(p^{-1},q))_n}{(p^{-1}-q)^n} \quad \mbox{for } \quad n\geq
1, \quad \end{array} \right.
\end{eqnarray}
respectively, which exactly reproduce the $(p,q)$-numbers and
$(p,q)-$factorials introduced by Chakrabarty and Jagannathan\cite{ChakJagan}.

The other properties can be recovered similarly to those of section \ref{janasec}
replacing the parameter $p$ by $p^{-1}$.

The $({\cal R},p,q)-$derivative is also reduced to $(p^{-1},q)-$derivative. Indeed,
\begin{eqnarray}
 \partial_{{\cal R},p,q} &=& \partial_{p,q}\frac{p-q}{P-Q}\frac{1-PQ}{(p^{-1}-q)P}
\cr&=&\frac{1}{(p^{-1}-q)z}(P^{-1}-Q)\equiv \partial_{p^{-1},q}
\end{eqnarray}
obtained by a simple replacement of the dilatation operator   $P$ by $P^{-1}$.

The algebra ${\cal A}_{p^{-1},q}$, generated by $\{1,\; A,\; A^\dag,\;N\}$, associated with
 $(p,q)-$Chakrabarty and Jagannathan deformation satisfies the following commutation relations:
\begin{eqnarray}
 A\;A^\dag- p^{-1}A^\dag A= q^N, \quad &&A\;A^\dag- qA^\dag A= p^{-N}\cr
[N,\;A^\dag]= A^\dag\qquad\qquad\quad&& [N,\;A]= -A.\label{ChakJagalg}
\end{eqnarray}


Hence, the
$(p^{-1},q)-$Rogers-Szeg\"o polynomials
\begin{eqnarray}
 H_n(z;p^{-1},q)=\sum_{k=0}^{n}\left[\begin{array}{c}n\\k\end{array}\right]_{p^{-1},q}z^k\quad n= 0, 1, 2, \cdots
\end{eqnarray}
obey the three-term recurrence relation
\begin{eqnarray}\label{ChakJagRogersttr}
H_{n+1}(z;p^{-1},q)&=& H_n(p^{-1}z;p^{-1},q) + z p^{-n}H_n(pz;p^{-1},q)
\cr&&\qquad-z(p^{-n}-q^n)H_{n-1}(z;p^{-1},q)
\end{eqnarray}
and $(p^{-1},q)-$difference equation
\begin{eqnarray}\label{ChakJagRogerdiff}
\partial_{p^{-1},q}H_n(z;p,q)= [n]_{p^{-1},q}H_{n-1}(z;p,q).
\end{eqnarray}
Finally, the set of polynomials
\begin{eqnarray}
 \psi_n(z;p^{-1},q)= \frac{1}{\sqrt{[n]!_{p^{-1},q}}}H_n(z;p^{-1},q),\quad n=0, 1, 2, \cdots
\end{eqnarray}
forms a basis for a realization of the $(p^{-1},q)-$deformed harmonic oscillator and quantum algebra ${\cal A}_{p^{-1},q}$
 generating
the commutation relations (\ref{ChakJagalg}) with the number
operator $N$ formally defined  as
\begin{eqnarray}
 N\psi_n(z;p^{-1},q)= n\psi_n(z;p^{-1},q),
\end{eqnarray}
and the annihilation and  creation operators  given by
\begin{eqnarray}
 A= \partial_{p^{-1},q}\quad\mbox{and}\quad A^\dag= P^{-1}+ zp^{-N}P-z(p^{-1}-q)\partial_{p^{-1},q},
\end{eqnarray}
respectively. Naturally, setting $p=1$ permits to recover the results of the subsection~\ref{qHosc}.

\subsection{${\cal R}(x,y)= \frac{xy-1}{(q-p^{-1})y}$}

In this case, the $({\cal R},p,q)-$numbers and $({\cal R}, p,q)-$factorials are reduced to 
\begin{eqnarray*}
 [n]_{p,q}^Q=\frac{p^n-q^{-n}}{q-p^{-1}},
\end{eqnarray*}
and
\begin{eqnarray}
[n]!_{p,q}^Q= \left\{\begin{array}{lr} 1 \quad \mbox{for   } \quad n=0 \quad \\
\frac{((p,q^{-1});(p,q^{-1}))_n}{(q-p^{-1})^n} \quad \mbox{for } \quad n\geq
1, \quad \end{array} \right.
\end{eqnarray}
introduced in our previous work \cite{HmNe},  generalizing the $q-$Quesne algebra \cite{QPT}.

Then follow some remarkable properties:
\begin{proposition}\label{PropoQes1}
 If $n$ and $m$ are nonnegative integers, then
\begin{eqnarray}
\;[-m]_{p,q}^Q&=& -p^{-m}q^m[m]_{p,q}^Q,\label{Qeq1}\\
\;[n+m]_{p,q}^Q&=& q^{-m}[n]_{p,q}^Q+p^n[m]_{p,q}^Q= p^m[n]_{p,q}^Q+q^{-n}[m]_{p,q}^Q,\label{Qeq2}\\
\;[n-m]_{p,q}^Q&=& q^{m}[n]_{p,q}^Q-p^{n-m}q^m[m]_{p,q}^Q= p^{-m}[n]_{p,q}^Q+p^{-m}q^{m-n}[m]_{p,q}^Q,\label{Qeq3}\\
\;[n]_{p,q}^Q &=& \frac{q-p^{-1}}{p-q^{-1}}[2]_{p,q}^Q[n-1]_{p,q}^Q-pq^{-1}[n-2]_{p,q}^Q.\label{Qeq4}
\end{eqnarray}
\end{proposition}
{\bf Proof:}  Eqs.(\ref{Qeq1}) and (\ref{Qeq2}) are immediate by the application of the relations
$ p^{-m}-q^{m} = -p^{-m}q^{m}(p^m-q^{-m})$ and
$p^{n+m}-q^{-n-m}= q^{-m}(p^{n}-q^{-n})+p^n(p^{m}-q^{-m})=p^m(p^{n}-q^{-n})+q^{-n}(p^{m}-q^{-m}),$
respectively, while Eq.(\ref{Qeq3}) results from the combination of Eqs.(\ref{Qeq1}) and (\ref{Qeq2}). Finally, the relation
\begin{eqnarray}\label{Qeq5}
[n]_{p,q^{-1}}&=& \frac{p^n-q^{-n}}{p-q^{-1}}= \frac{q-p^{-1}}{p-q^{-1}}\frac{p^n-q^{-n}}{q-p^{-1}}
= \frac{q-p^{-1}}{p-q^{-1}}[n]_{p,q}^{Q},\; n=1, 2, \cdots
\end{eqnarray}
cumulatively taken with the  identity
\begin{eqnarray*}
 [n]_{p,q{-1}}&=& [2]_{p,q{-1}}[n-1]_{p,q{-1}}-pq^{-1}[n-2]_{p,q^{-1}}
\end{eqnarray*}
gives Eq.(\ref{Qeq4}).\hfill$\Box$

\begin{proposition}\label{PropoQes2}
The $(p,q)-$Quesne binomial coefficients
\begin{eqnarray}
 \left[\begin{array}{c} n \\ k \end{array}\right]_{p,q}^Q=
\frac{((p,q^{-1});(p,q^{-1}))_n}{((p,q^{-1});(p,q^{-1}))_k((p,q^{-1});(p,q^{-1}))_{n-k}}, \label{Qeq6}
\end{eqnarray}
where $\quad 0\leq k\leq n;\;\; n\in\mathbb{N},$ satisfy the following properties
\begin{eqnarray}
\left[\begin{array}{c} n \\ k \end{array}\right]_{p,q}^Q
= \left[\begin{array}{c} n \\ n-k \end{array}\right]_{p,q}^Q=
p^{k(n-k)}\left[\begin{array}{c} n \\ k \end{array}\right]_{1/qp}=
p^{k(n-k)}\left[\begin{array}{c} n \\ n-k \end{array}\right]_{1/qp},\label{Qeq7}
\end{eqnarray}
\begin{eqnarray}
\left[\begin{array}{c} n+1 \\ k \end{array}\right]_{p,q}^Q =
p^k\left[\begin{array}{c} n \\ k \end{array}\right]_{p,q}^Q
+q^{-n-1+k}\left[\begin{array}{c} n \\ k-1 \end{array}\right]_{p,q}^Q,\label{Qeq8}
\end{eqnarray}
\begin{eqnarray}
 \left[\begin{array}{c} n+1 \\ k \end{array}\right]_{p,q}^Q &=&
p^{k}\left[\begin{array}{c} n \\ k \end{array}\right]_{p,q}^Q +
p^{n+1-k}\left[\begin{array}{c} n \\ k-1 \end{array}\right]_{p,q}^Q 
\cr&&\qquad\quad-(p^n-q^{-n})
\left[\begin{array}{c} n-1 \\ k-1 \end{array}\right]_{p,q}^Q.\quad\label{Qeq9}
\end{eqnarray}
\end{proposition}
{\bf Proof:} It is straightforward, using the Proposition~\ref{PropJan1}  and
\begin{eqnarray}
 \left[\begin{array}{c} n \\ k \end{array}\right]_{p,q}^Q =
\left[\begin{array}{c} n \\ k \end{array}\right]_{p,q^{-1}}.
\end{eqnarray}
\hfill$\Box$\\
Finally, the algebra ${\cal A}_{p,q}^Q$, generated by $\{1,\; A,\; A^\dag,\;N\}$, associated with
 $(p,q)-$Quesne deformation satisfies the following commutation relations:
\begin{eqnarray}
 p^{-1}A\;A^\dag- A^\dag A= q^{-N-1}, \quad&& qA\;A^\dag- A^\dag A= p^{N+1}\cr
[N,\;A^\dag]= A^\dag,\qquad\qquad\qquad&& [N,\;A]= -A.\label{Qalg}
\end{eqnarray}
The ($p,q)-$Rogers-Szeg\"o polynomials corresponding to the Quesne
deformation\cite{HmNe} are deduced from our generalization by choosing
$\phi_1(x,y)=\phi_2(x,y)=\phi(x,y)= x$ and $\phi_3(x,y)= y-x^{-1}$. Indeed, it is worthy of attention that we get in this case
$\phi(p,q)= p\neq 0$, $ \phi_3(p,q)=q-p^{-1}\neq 0$, $\phi(P,Q)z^k= \phi_1^k(p,q)z^k$ and from Eq.(\ref{Qeq9})
$$\left[\begin{array}{c} n+1 \\ k \end{array}\right]_{p,q}^Q =
p^{k}\left[\begin{array}{c} n \\ k \end{array}\right]_{p,q}^Q +
p^{n+1-k}\left[\begin{array}{c} n \\ k-1 \end{array}\right]_{p,q}^Q -(q-p^{-1})[n]_{p,q}
\left[\begin{array}{c} n-1 \\ k-1 \end{array}\right]_{p,q}^Q.$$
Hence, the hypotheses of the theorem are satisfied and, therefore,
the $(p,q)-$Rogers-Szeg\"o polynomials
\begin{eqnarray}
 H_n^Q(z;p,q)=\sum_{k=0}^{n}\left[\begin{array}{c}n\\k\end{array}\right]_{p,q}^Qz^k,\quad n= 0, 1, 2, \cdots
\end{eqnarray}
satisfy the three-term recurrence relation
\begin{eqnarray}\label{QRogersttr}
H_{n+1}^Q(z;p,q)= H_n^Q(pz;p,q) &+& z p^nH_n^Q(p^{-1}z;p,q)
\cr&-&z(p^n-q^{-n})H_{n-1}^Q(z;p,q)
\end{eqnarray}
and the $(p,q)-$difference equation
\begin{eqnarray}\label{QRogerdiff}
\partial_{p,q}^QH_n^Q(z;p,q)= [n]_{p,q}^QH_{n-1}^Q(z;p,q).
\end{eqnarray}
Thus, the set of polynomials
\begin{eqnarray}
 \psi_n^Q(z;p,q)= \frac{1}{\sqrt{[n]!_{p,q}^Q}}H_n^Q(z;p,q),\quad n=0, 1, 2, \cdots
\end{eqnarray}
forms a basis for a realization of the $(p,q)-$ Quesne deformed
harmonic oscillator and quantum algebra ${\cal A}_{p,q}^Q$
engendering the commutation relations (\ref{Qalg}) with the number
operator $N$  formally defined  as
\begin{eqnarray}
 N\psi_n^Q(z;p,q)= n\psi_n^Q(z;p,q),
\end{eqnarray}
and the annihilation and  creation operators  given by
\begin{eqnarray}
 A= \partial_{p,q}^Q\quad\mbox{and}\quad A^\dag= P+ zp^NP^{-1}-z(q-p^{-1})\partial_{p,q},
\end{eqnarray}
respectively. Naturally, setting $p=1$ gives the Rogers-Szeg\"o
polynomials associated with the $q-$Quesne deformation~\cite{QPT}.

The continuous $(p,q)-$Hermite polynomials corresponding to the $(p, q)-$ generalization of  Quesne deformation\cite{HmNe} can be defined
as follows:
\begin{eqnarray}
 \mathbb{H}_n^Q(\cos\theta;p,q)&=& e^{in\theta}H_n^Q(e^{-2i\theta};p,q)
\cr&=&
\sum_{k=0}^n\left[\begin{array}{c}n\\k\end{array}\right]_{p,q}^Q\;e^{i(n-2k)\theta},\quad n=0, 1, 2, \cdots.
\end{eqnarray}
Since for the $(p, q)$-generalization of  Quesne deformation \cite{HmNe} $\phi_1(x,y)=\phi_2(x,y)= x$ and
$\phi_3(x,y)=y-x^{-1}$, from the Proposition \ref{propBuk} we
deduce that the corresponding sequence of continuous
$(p,q)-$Hermite polynomials satisfies the three-term recurrence
relation
\begin{eqnarray}\label{QHttr}
 \mathbb{H}_{n+1}^Q(\cos\theta;p,q)&=&p^{\frac{n}{2}}(e^{i\theta}P+e^{-i\theta}P^{-1})\mathbb{H}_n^Q(\cos\theta;p,q)
\cr&&\qquad-
(p^n-q^{-n})\mathbb{H}_{n-1}^Q(\cos\theta;p,q).
\end{eqnarray}


\subsection{$\displaystyle{\cal R}(x,y)= h(p,q)y^\nu/x^\mu\left[\frac{xy-1}{(q-p^{-1})y}\right]$}
%
Here
$0< pq < 1$ , $p^\mu<q^{\nu-1}$, $p>1$ ,  $h$ is a well behaved
real and non-negative function of deformation parameters $p$ and
$q$ such that  $h(p,q)\to 1$ as $(p,q)\to (1,1).$

The $({\cal R},p,q)-$numbers become $(p,q;\mu,\nu,h)$-numbers introduced in our previous work \cite{HmNe2} and defined by
\begin{eqnarray}
 [n]^{\mu,\nu}_{p,q,h}= h(p,q)\frac{q^{\nu n}}{p^{\mu n}}\frac{p^n-q^{-n}}{q-p^{-1}}.
\end{eqnarray}
\begin{proposition}\label{PropoHouk1}
The $(p,q;\mu,\nu,h)-$numbers verify the following properties, for $m,n\in\mathbb{N}$:
\begin{eqnarray}
\;[-m]^{\mu,\nu}_{p,q,h}= -\frac{q^{-2\nu m+m}}{p^{-2\mu m+m}}[m]^{\mu,\nu}_{p,q,h},\label{Heq1}
\end{eqnarray}
\begin{eqnarray}
\;[n+m]^{\mu,\nu}_{p,q,h}&=& \frac{q^{\nu m-m}}{p^{\mu m}}[n]^{\mu,\nu}_{p,q,h}
+\frac{q^{\nu n}}{p^{\mu n-n}}[m]^{\mu,\nu}_{p,q,h}
\cr&=&\frac{q^{\nu m}}{p^{\mu m- m}}[n]^{\mu,\nu}_{p,q,h}
+\frac{q^{\nu n- n}}{p^{\mu n}}[m]^{\mu,\nu}_{p,q,h}\;,\label{Heq2}
\end{eqnarray}
\begin{eqnarray}
\;[n-m]^{\mu,\nu}_{p,q,h}&=&\frac{q^{-\nu m+ m}}{p^{-\mu m}}[n]^{\mu,\nu}_{p,q,h}
-\frac{q^{\nu(n-2m)+m}}{p^{\mu(n-2m)-n+m}}[m]^{\mu,\nu}_{p,q,h}
\cr&=& \frac{q^{-\nu m}}{p^{-\mu m+m}}[n]^{\mu,\nu}_{p,q,h}
-\frac{q^{\nu(n-2m)-n+m}}{p^{\mu(n-2m)+m}}[m]^{\mu,\nu}_{p,q,h},\label{Heq3}
\end{eqnarray}
\begin{eqnarray}
\; [n]^{\mu,\nu}_{p,q,h}= \frac{q-p^{-1}}{p-q^{-1}}\frac{q^{-\nu}}{p^{-\mu}}\frac{1}{h(p,q)}[2]^{\mu,\nu}_{p,q,h}
[n-1]^{\mu,\nu}_{p,q,h}-\frac{q^{2\nu-1}}{p^{2\nu-1}}[n-2]^{\mu,\nu}_{p,q,h}.\label{Heq4}
\end{eqnarray}
\end{proposition}
{\bf Proof:} It is direct using the Proposition~\ref{PropoQes1} and the fact that
\begin{eqnarray}\label{Houkeq}
 [n]^{\mu,\nu}_{p,q,h}= h(p,q)\frac{q^{\nu n}}{p^{\mu n}}[n]^Q_{p,q}.
\end{eqnarray}
\hfill$\Box$
\begin{proposition}\label{PropoHouk2}
The $(p,q,\mu,\nu,h)-$ binomial coefficients
\begin{eqnarray}
 \left[\begin{array}{c} n \\ k \end{array}\right]_{p,q,h}^{\mu,\nu}:=
\frac{[n]!_{p,q,h}^{\mu,\nu}}{[k]!_{p,q,h}^{\mu,\nu}[n-k]!_{p,q,h}^{\mu,\nu}}=
\frac{q^{\nu k(n-k)}}{p^{\mu k(n-k)}}\left[\begin{array}{c} n \\ k \end{array}\right]_{p,q}^Q,
\label{Heq5}
\end{eqnarray}
where $\;0\leq k\leq n;\;\; n\in\mathbb{N},$ satisfy the following properties
\begin{eqnarray}
\left[\begin{array}{c} n \\ k \end{array}\right]_{p,q,h}^{\mu,\nu}
= \left[\begin{array}{c} n \\ n-k \end{array}\right]_{p,q,h}^{\mu,\nu},\label{Heq6}
\end{eqnarray}
\begin{eqnarray}
 \left[\begin{array}{c} n+1 \\ k \end{array}\right]_{p,q,h}^{\mu,\nu} =
\frac{q^{\nu k}}{p^{(\mu-1)k}}\left[\begin{array}{c} n \\ k \end{array}\right]_{p,q,h}^{\mu,\nu}
+\frac{q^{(\nu-1)(n+1-k)}}{p^{\mu(n+1-k)}}\left[\begin{array}{c} n \\ k-1 \end{array}\right]_{p,q,h}^{\mu,\nu},
\label{Heq7}
\end{eqnarray}
\begin{eqnarray}
 \left[\begin{array}{c} n+1 \\ k \end{array}\right]_{p,q,h}^{\mu,\nu} &=&
\frac{q^{\nu k}}{p^{(\mu-1)k}}\left[\begin{array}{c} n \\ k \end{array}\right]_{p,q,h}^{\mu,\nu} +
\frac{q^{\nu(n+1-k)}}{p^{(\mu-1)(n+1-k)}}\left[\begin{array}{c} n \\ k-1 \end{array}\right]_{p,q,h}^{\mu,\nu}
\cr&&\qquad\qquad\quad\quad-(p^n-q^{-n})
\frac{q^{\nu n}}{p^{\mu n}}\left[\begin{array}{c} n-1 \\ k-1 \end{array}\right]_{p,q,h}^{\mu,\nu}.\quad\label{Heq8}
\end{eqnarray}
\end{proposition}
{\bf Proof:} There follow from the Proposition~\ref{PropoQes2} and the fact that
\begin{eqnarray}
 [n]!_{p,q,h}^{\mu,\nu}= h^n(p,q)\frac{q^{n(n+1)/2}}{p^{n(n+1)/2}}[n]!_{p,q}^Q,
\end{eqnarray}
where  use of Eq.(\ref{Houkeq}) has been made.\hfill$\Box$\\
The algebra ${\cal A}_{p,q,h}^{\mu,\nu}$, generated by
$\{1,\; A,\; A^\dag,\;N\}$, associated with
 $(p,q,\mu,\nu,h)$-deformation, satisfies the following commutation relations:
\begin{eqnarray}
&& p^{-1}A\;A^\dag- \frac{q^\nu}{p^\mu} A^\dag A= h(p,q)\left(\frac{q^{\nu-1}}{p^\mu}\right)^{N+1},
\cr&&
qA\;A^\dag- \frac{q^\nu}{p^\mu}A^\dag A= h(p,q)\left(\frac{q^\nu}{p^{\mu-1}}\right)^{N+1}\qquad\cr
&&[N,\;A^\dag]= A^\dag,\qquad\qquad [N,\;A]= -A.\label{Hkalg}
\end{eqnarray}

The $(p,q,\mu,\nu,h)$-Rogers-Szeg\"o\cite{HmNe2} polynomials are deduced from
the above general construction  by setting $\phi_1(x,y)= x^{1-\mu}y^\nu$,
$\phi_2(x,y) = x^{-\mu}y^{\nu-1}$ and $\phi_3(x,y)=
\frac{y-x^{-1}}{h(p,q)}$. Indeed, $\phi_i(p,q)\neq 0$ for $i=1, 2,
3$; $\phi_i(P,Q)z^k = \phi_i(p,q)^kz^k$ for $i= 1, 2$ and  the
property (\ref{Heq8}) furnishes
\begin{eqnarray*}
\left[\begin{array}{c} n+1 \\ k \end{array}\right]_{p,q,h}^{\mu,\nu} &=&
\frac{q^{\nu k}}{p^{(\mu-1)k}}\left[\begin{array}{c} n \\ k \end{array}\right]_{p,q,h}^{\mu,\nu} +
\frac{q^{\nu(n+1-k)}}{p^{(\mu-1)(n+1-k)}}\left[\begin{array}{c} n \\ k-1 \end{array}\right]_{p,q,h}^{\mu,\nu}
\cr&&\qquad-
\frac{q-p^{-1}}{h(p,q)}[n]^{\mu,\nu}_{p,q,h}\left[\begin{array}{c} n-1 \\ k-1 \end{array}\right]_{p,q,h}^{\mu,\nu}.
\end{eqnarray*}
Therefore, the $(p,q,\mu,\nu,h)$-Rogers-Szeg\"o polynomials are defined as follows:
\begin{eqnarray}
H_n(z;p,q,\mu,\nu,h)=\sum_{k=0}^{n}\left[\begin{array}{c}n\\k\end{array}\right]_{p,q,h}^{\mu,\nu} z^k,\quad n=0,1,2\cdots
\end{eqnarray}
with the three-term recurrence relation
\begin{eqnarray}\label{HkRogersttr}
H_{n+1}(z;p,q,\mu,\nu,h)&=& H_n\left(\frac{q^{\nu}}{p^{\mu-1}}z:p,q,\mu,\nu,h\right)
\cr&&\qquad+ z\frac{q^{(\nu-1)n}}{p^{\mu n}}H_n\left(\frac{p^{\nu}}{q^{\nu-1}}z;p,q,\mu,\nu,h\right)
\cr&&\qquad-z\frac{q^{\nu n}}{p^{\mu n}}(p^n-q^{-n})H_{n-1}(z;p,q,\mu,\nu,h)
\end{eqnarray}
and $(p,q,\mu,\nu,h)-$difference equation
\begin{eqnarray}\label{HkRogerdiff}
\partial_{p,q,h}^{\mu,\nu}H_n(z;p,q,\mu,\nu,h)= [n]_{p,q,h}^{\mu,\nu}H_{n-1}(z;p,q,\mu,\nu,h).
\end{eqnarray}
Hence, the set of polynomials
\begin{eqnarray}
 \psi_n(z;p,q,\mu,\nu,h)= \frac{1}{\sqrt{[n]!_{p,q,h}^{\mu,\nu}}}H_n(z;p,q,\mu,\nu,h),\quad n=0, 1, 2, \cdots
\end{eqnarray}
forms a basis for a realization of the $(p,q,\mu,\nu,h)-$deformed algebra ${\cal A}_{p,q,\mu,\nu,h}$ satisfying the commutation relations (\ref{Hkalg}) with the number
operator $N$ formally defined  as
\begin{eqnarray}
 N\psi_n^Q(z;p,q,\mu,\nu,h)= n\psi_n(z;p,q,\mu,\nu,h),
\end{eqnarray}
together with the annihilation and the creation operators  given by
\begin{eqnarray}
 A= \partial_{p,q,h}^{\mu,\nu}\;\mbox{ and }\;A^\dag= \frac{Q^{\nu}}{P^{\mu-1}}
+ z\left(\frac{q^{\nu-1}}{p^{\mu}}\right)^N\frac{P^{\mu}}{Q^{\nu-1}}-z\frac{(q-p^{-1})}{h(p,q)}\partial_{p,q,h}^{\mu,\nu},
\end{eqnarray}
respectively.

The continuous $(p,q,\mu,\nu,h)-$Hermite polynomials\cite{HmNe2} can be now deduced as:
\begin{eqnarray}
 \mathbb{H}_n(\cos\theta;p,q,\mu,\nu,h)&=& e^{in\theta}H_n(e^{-2i\theta};p,q,\mu,\nu,h)\cr&=&
\sum_{k=0}^n\left[\begin{array}{c}n\\k\end{array}\right]_{p,q,h}^{\mu,\nu}\;e^{i(n-2k)\theta},\quad n=0, 1, 2, \cdots.
\end{eqnarray}
Since for the $(p,q,\mu,\nu,h)-$deformation
$\phi_1(x,y)= x^{1-\mu}y^\nu$, $\phi_2(x,y) = x^{-\mu}y^{\nu-1}$ and $\phi_3(x,y)= \frac{y-x^{-1}}{h(p,q)}$,
from the Proposition \ref{propBuk}  the corresponding sequence of continuous
$(p,q,\mu,\nu,h)-$Hermite polynomials satisfies the three-term recurrence relation
\begin{eqnarray}\label{HkHttr}
 \mathbb{H}_{n+1}(\cos\theta;p,q,\mu,\nu,h)&=&
\frac{q^{\nu\frac{n}{2}}}{p^{(\mu-1)\frac{n}{2}}}
\frac{Q^{\nu}}{P^{\mu-1}}\mathbb{H}_n(\cos\theta;p,q,\mu,\nu,h)\cr&&
+\frac{q^{(\nu-1)\frac{n}{2}}}{p^{\mu\frac{n}{2}}}
\frac{Q^{-(\nu-1)}}{P^{-\mu}}
\mathbb{H}_n(\cos\theta;p,q,\mu,\nu,h)\cr&&
-(p^n-q^{-n})\frac{q^{\nu n}}{p^{\mu n}}\;\mathbb{H}_{n-1}(\cos\theta;p,q,\mu,\nu,h).
\end{eqnarray}

\section{Concluding remarks}\label{Sec6}
In this paper, we have defined and discussed  a general formalism for constructing $({\cal R},p,q)-$ deformed Rogers-Szeg\"o
polynomials.   The displayed approach not only provides novel relations, but also generalizes  well known standard and deformed
 Rogers-Szeg\"o polynomials. A full characterization of the latter, including the data on the 
three-term recurrence relations and
difference equations, has been provided.
 We have succeeded in elaborating  a new realization of
 $({\cal R},p,q)-$deformed quantum algebra generalizing the construction of 
$q-$deformed harmonic oscillator creation and annihilation
operators  performed in
 \cite{Galetti,JaganSridh}. The  continuous $({\cal R},p,q)-$Hermite polynomials have been also investigated in detail.

Finally, relevant particular cases and examples have been exhibited.


\section*{Acknowledgements}
This work is partially supported by the Abdus Salam International
Centre for Theoretical Physics (ICTP, Trieste, Italy) through the
Office of External Activities (OEA) - \mbox{Prj-15}. The ICMPA
is in partnership with
the Daniel Iagolnitzer Foundation (DIF), France.

\end{document}